\documentclass[9pt,conference]{IEEEtran}
\usepackage{waspaa25} % removes hyperlinks (required by IEEE Xplore)

\usepackage{amsmath} % For mathematical symbols and environments
\usepackage{amssymb} % For additional math symbols
\usepackage{multirow}
\DeclareMathOperator*{\argminA}{arg\,min}
\newcommand{\norm}[1]{\left\lVert#1\right\rVert}
	
\usepackage{bm} % for bold math symbols (incl. Greek letters) with \bm{}

\title{Latent Acoustic Mapping for Direction of Arrival Estimation: A Self-Supervised Approach}

\name{Adrian S. Roman$^{1}$,
      Iran R. Roman$^{2}$,
      Juan P. Bello$^{3}$,
      }
\address{$^{1}$University of Southern California, Los Angeles, USA \; $^{2}$Queen Mary University of London, London, UK \\$^{3}$New York University, New York, USA
}

\begin{document}

\maketitle

\begin{abstract}
Acoustic mapping techniques have long been used in spatial audio processing for direction of arrival estimation (DoAE). Traditional beamforming methods for acoustic mapping, while interpretable, often rely on iterative solvers that can be computationally intensive and sensitive to acoustic variability. On the other hand, recent supervised deep learning approaches offer feedforward speed and robustness but require large labeled datasets and lack interpretability. Despite their strengths, both methods struggle to consistently generalize across diverse acoustic setups and array configurations, limiting their broader applicability. We introduce the Latent Acoustic Mapping (LAM) model, a self-supervised framework that bridges the interpretability of traditional methods with the adaptability and efficiency of deep learning methods. LAM generates high-resolution acoustic maps, adapts to varying acoustic conditions, and operates efficiently across different microphone arrays. We assess its robustness on DoAE using the LOCATA and STARSS benchmarks. LAM achieves comparable or superior localization performance to existing supervised methods. Additionally, we show that LAM’s acoustic maps can serve as effective features for supervised models, further enhancing DoAE accuracy and underscoring its potential to advance adaptive, high-performance sound localization systems.
\end{abstract}

\section{Introduction}
\label{sec:intro}

Acoustic mapping techniques have long played a central role in spatial audio processing, particularly for sound direction of arrival estimation (DoAE). Also referred to as acoustic imaging, acoustic mapping offers a spatial visualization of sound sources, facilitating intuitive tracking of sound emissions in complex environments \cite{brooks2006deconvolution,simeoni2019deepwave}.

Traditional acoustic mapping methods have laid the foundation for interpretable spatial sound analysis \cite{bryn1962optimum,schmidt1986multiple,brooks2006deconvolution}. These typically project a directional sound intensity field onto a spatial representation by steering the array’s sensitivity pattern across directions of interest \cite{krim1996two, schmidt1986multiple}. While computationally efficient, the resulting acoustic images suffer from poor angular resolution due to diffraction limits imposed by array size and sound wavelength \cite{simeoni2019deepwave,chardon2021theoretical}. Newer compressed sensing methods promise higher fidelity \cite{simeoni2019deepwave}, but their reliance on iterative solvers makes them computationally prohibitive \cite{yardibi2008sparsity, chardon2021theoretical}.

Deep learning has emerged as a promising alternative for acoustic mapping, offering improved resolution without the computational burden of iterative solvers. DeepWave \cite{simeoni2019deepwave} exemplifies this shift by integrating graph-based convolutions within a recurrent architecture to produce high-resolution acoustic maps efficiently. It bridges traditional spatial filtering with data-driven learning, enabling practical deployment in real-time settings. However, DeepWave depends on ground-truth labels generated by compressed sensing methods, which constrain its performance ceiling. Its supervised nature and reliance on acoustic maps raise concerns about overfitting and generalization \cite{roman2024robust}, prompting growing interest in end-to-end models that learn directly from raw multichannel audio.

\begin{figure*}
\centering
\includegraphics[width=0.9\textwidth]{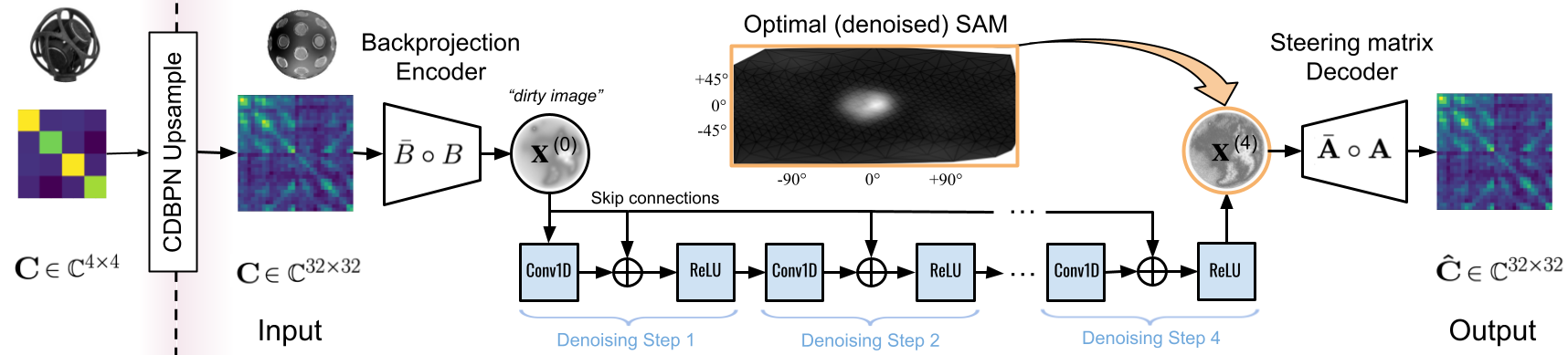}
\caption{LAM Architecture: LAM encodes and reconstructs microphone covariance matrix via self-supervised learning. The latent SAM is denoised using 4 denoising steps each with 1D convolutions followed by ReLU non-linearities. The optimal (denoised) SAM $\mathbf{x}^{(4)}$ is then decoded, guided by the steering matrix, to estimate the input CSM. Super-resolution upsampling (CDBPN) enables processing of inputs from lower-resolution arrays.}
\label{fig:LAM}
\end{figure*}

Building on this body of research, we propose the \textbf{L}atent \textbf{A}coustic \textbf{M}apping (LAM) model, a self-supervised model to generate high-resolution spherical acoustic maps (SAMs) from both low- and high-resolution microphone arrays. Figure \ref{fig:LAM} shows LAM's architecture. LAM synthesizes insights from traditional acoustic mapping, signal processing and self-supervised data-driven learning, offering scalability and adaptability to large and diverse datasets. It employs an autoencoder architecture to encode the covariance matrix of a microphone array into a latent space that captures spatially localized energy patterns in an interpretable acoustic map. Inspired by the graph convolutions from DeepWave \cite{simeoni2019deepwave} and recent image denoising techniques \cite{zhang2017beyond, ilesanmi2021methods}, LAM further refines the acoustic map by applying denoising convolutions. LAM’s self-supervised approach leverages a decoder to reconstruct the microphone channel correlation matrix using the array's \textit{steering matrix} \cite{krim1996two}. Unlike DeepWave, which inherits the limitations of its supervised training labels, LAM learns directly from raw multichannel data, removing the ceiling imposed by iterative-method-derived ground truth. Additionally, LAM has an upsampling module, enabling its adaptation to different microphone configurations, bridging the gap between low- and high-resolution covariance matrices.

\begin{figure}
% \centering
\resizebox{\columnwidth}{!}{
\includegraphics{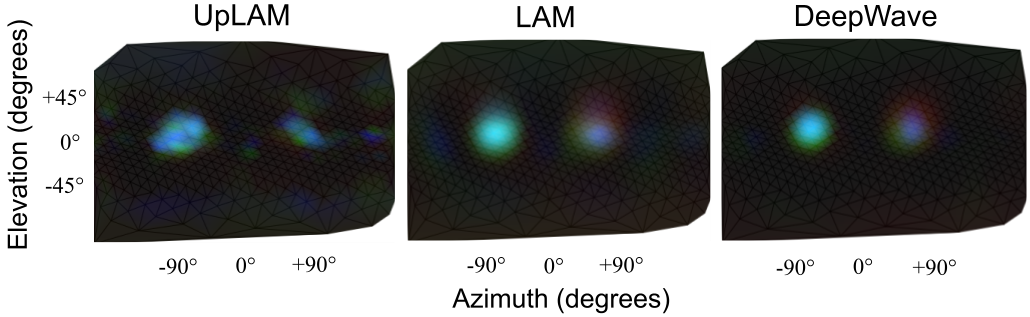}}
\caption{Spherical Acoustic Maps (SAMs) for two active speech sound sources generated using UpLAM (4ch), LAM (32ch) and DeepWave (32ch) \cite{simeoni2019deepwave}.}
\label{fig:intensity_maps}
\end{figure}

We evaluate LAM on datasets for DoAE, demonstrating its ability to outperform or match existing supervised methods while being robust across varied acoustic conditions\footnote{\href{https://github.com/adrianSRoman/LAM}{\texttt{https://github.com/adrianSRoman/LAM}}}. Our key contributions include:

\begin{itemize} 
    \item LAM: A self-supervised model for generating high-resolution acoustic maps from microphone array recordings.

    \item A benchmark showcasing LAM’s superior DoAE performance compared to DeepWave and signal processing methods for acoustic mapping. % SAM generation. %geDeepWave on the LOCATA speech localization dataset.

    \item An analysis of LAM’s adaptability and robustness across low- and high-resolution microphone arrays.
\end{itemize}

\section{Related work}
\label{sec:related}

\subsection{Signal-processing methods for acoustic mapping}

The fundamental method for acoustic mapping is delay-and-sum (DAS) beamforming. 
It linearly combines correlated microphone signals to steer the array’s sensitivity pattern across directions of interest, producing a spatial heatmap of acoustic energy \cite{bryn1962optimum}. While computationally efficient and interpretable, DAS suffers from poor angular resolution due to diffraction limits \cite{schmidt1986multiple}. In practical terms, this results in coarse directional estimates, especially at low frequencies or when using compact arrays typical in mobile or embedded applications. The MUltiple SIgnal Classification (MUSIC) algorithm addresses some of these limitations by leveraging the eigenstructure of the spatial covariance matrix to separate signal and noise subspaces, enabling super-resolution localization beyond the diffraction limit \cite{schmidt1986multiple}. However, MUSIC relies on assumptions such as the number of sound sources (to draw a separation between signals and noise) and them being uncorrelated. It also becomes less robust in noisy conditions. 

Compressed sensing approaches have been proposed to address these limitations \cite{simeoni2019deepwave}. Methods like the  deconvolution approach for the mapping of acoustic sources (DAMAS) \cite{brooks2006deconvolution} and variants of Proximal Gradient Descent (PGD) \cite{simeoni2019deepwave, beck2009fast} resolve finer angular details by imposing sparsity. However, these depend on iterative solvers with non-trivial convergence times. Thus, many applications such as acoustic cameras still rely on DAS due to its low-latency performance.

LAM addresses these limitations by learning spatial energy distributions directly from the microphone covariance matrix, thus bypassing the diffraction-limited beam patterns of DAS and the eigen subspace separation assumptions required by MUSIC. It also  eliminates the need for iterative solvers, instead using a feedforward architecture.% that generates high-resolution acoustic maps.

\subsection{Deep acoustic mapping}

Data-driven and deep learning approaches have emerged. DeepWave \cite{simeoni2019deepwave} was a first-of-its-kind system that leveraged supervised learning to approximate high-resolution SAMs traditionally obtained through iterative solvers \cite{gregor2010learning}. By building on these methods, DeepWave achieved high resolution acoustic imaging while maintaining computational efficiency inherited from feedforward networks. Previous work has also shown that the SAMs generated by DeepWave can serve as robust features for DoAE tasks \cite{roman2024robust}. However, its supervised nature presents generalization challenges as its learning can be bounded by pre-computed labels that are in turn generated from iterative solvers.
In contrast, LAM uses a self-supervised learning approach, enabling learning from raw multichannel audio and no pre-computed labels. This improves generalization and removes performance ceilings.

\subsection{Neural Networks for DoAE}

Neural networks for sound event localization and detection (SELD) have demonstrated strong DoAE performance on benchmarks such as LOCATA and STARSS \cite{adavanne2019localization, hu2022track, Shimada2023starss23,kushwaha2023sound}. These approaches are predominantly supervised, leveraging datasets annotated with activity and direction-of-arrival labels \cite{grumiaux2022survey}. However, due to the complexity and cost of generating these annotations, SELD models are often optimized for specific microphone array configurations and a limited range of acoustic environments. While this results in high accuracy for various benchmarks, it restricts the generalization of these models to real-world scenarios where microphone configurations and training data can vary significantly~\cite{hu2022track, politis2020overview}.
LAM's self-supervised learning allows for use of unlabeled data at scale, and its microphone upsampling module allows it to adapt to varying array configurations.

\section{Self-Supervised LAM model}
\label{sec:approach}

\subsection{Definitions}

As it is common in compressed sensing\cite{johnson1992array}, to compute a SAM we operate in the frequency domain and rely on the computation of the Cross Spectral Matrix (CSM) and a known steering matrix.

The CSM captures the spatial covariance of signals across the microphone channels. To compute a CSM, consider a multichannel audio recording $\mathbf{y} \in \mathbb{R}^{M\times T}$ with $M$ microphone channels and duration $T$. The Short-time Fourier Transform (STFT) of $\mathbf{y}$, denoted as $\mathbf{Y}(n, k) \in \mathbb{C}^{M}$, represents the complex-valued frequency-domain representation at time frame $n$ and frequency band $k$. Given $N$ subsequent time frames, the CSM at the $k\text{th}$ band is defined as:
\begin{equation}
\mathbf{C}_{k} = \frac{1}{N} \sum_{n=1}^{N} \mathbf{Y}(n,k)\mathbf{Y}(n,k)^H  \in \mathbb{C}^{M\times M},
\label{eq:cov}
\end{equation}
where $\mathbf{Y}(n,k)^H$ is the Hermitian transpose of $\mathbf{Y}(n,k)$. 

The steering matrix captures the phase shifts and delays needed to focus on a source at a specific location \cite{van2013signal}. It is defined as:
\begin{equation}
\mathbf{A} = \text{exp} (-j\frac{2 \pi}{\lambda_0} \mathbf{P}^T \mathbf{R}) \in \mathbb{C}^{M\times N},
\end{equation} 
which consists of the coordinates for $M$ microphone channels $\mathbf{P} = \{\mathbf{p_1}, ..., \mathbf{p_M}\} \in\mathbb{R}^{3 \times M}$, the spherical tessellation points $\mathbf{R} = \{\mathbf{r_1}, ..., \mathbf{r_N} \} \in\mathbb{R}^{3 \times N}$, $\lambda_0$ is the wavelength, and $j = \sqrt{-1}$. 

\subsection{Encoder}
\label{sec:encoder}

LAM takes as input a CSM $\mathbf{C}$ and
encodes it via the linear operation:
\begin{equation}
\mathbf{x}^{(0)} = [\bar{\textbf{B}}\circ\textbf{B}]^H \text{vec}(\mathbf{C}) \in \mathbb{R}^N, 
\label{eq:encode}
\end{equation}
where $\textbf{B} \in \mathbb{C}^{M\times N}$ is a learnable matrix. $\circ$ is the Khatri-Rao product \cite{liu2008hadamard} and vec: $\in \mathbb{C}^{M\times M} \rightarrow \mathbb{C}^{M^2}$ ``flattens'' matrix $\mathbf{C}$. Eq. \ref{eq:encode} is referred to as the ``back-projection,'' which maps $\mathbf{C}$ to the latent space, in our case structured on a Fibonacci tessellation~\cite{kushwaha2022analyzing}. The initial projection, $\mathbf{x}^{(0)}$, represents a \textit{dirty image} \cite{van2013signal} (a noisy SAM). LAM processes this projection through 4 denoising steps. Each step employs denoising 1D convolutions with progressively larger kernel sizes (3, 5, 7, and 9) \cite{simeoni2019deepwave, ilesanmi2021methods}. The layers also incorporate residual connections from $\mathbf{x}^{(0)}$, and apply ReLU non-linearities. The SAMs at each step are denoted as $\mathbf{x}^{(1)}$, $\mathbf{x}^{(2)}$, $\mathbf{x}^{(3)}$, and $\mathbf{x}^{(4)}$. The optimal (denoised) SAM is $\mathbf{x}^{(4)}$.

\subsection{Decoder}
\label{sec:decoder}

\noindent Given a steering matrix \(\mathbf{A}\), one can use the final \(\mathbf{x}^{(4)}\) to reconstruct the covariance matrix \(\mathbf{\hat{C}}\) as:

\begin{equation}
\mathbf{C} \approx \mathbf{\hat{C}} = \mathbf{A}\ \text{diag}(\mathbf{x}^{(4)})\ \mathbf{A}^{\text{H}} = (\bar{\mathbf{A}}\circ \mathbf{A})\mathbf{x}^{(4)},
\label{eq:steer}
\end{equation}
where \(\mathbf{\hat{C}}\) approximates the true CSM \(\mathbf{C}\), diag($\cdot$) transforms $\mathbf{x}^{(4)}$ into a diagonal $N\times N$ matrix.

\subsection{Self-supervised Pre-text Task}

Our goal is to learn %a high-resolution SAM 
\(\mathbf{x}^{(4)}\) %using the encoder described in Section \ref{sec:encoder}, enabling an 
to optimally estimate \(\mathbf{\hat{C}}\). We can self-supervise the estimation of $\mathbf{\hat{C}}$ via the optimization of:

% \begin{equation}
% \mathbf{x} \in \argminA_{\mathbf{x} \in \mathbb{R}^{N}_{\geq 0}} \| \mathbf{C} - \mathbf{A} \text{diag}(\mathbf{x}) \mathbf{A}^H\|_2^2, %+ \gamma \norm{\mathbf{x}}_1,
% \end{equation}

\begin{equation}
\mathbf{x}^{(4)} \in \argminA_{\mathbf{x}^{(4)} \in \mathbb{R}^{N}_{\geq 0}} \norm{\mathbf{C} - \mathbf{A} \, \text{diag}(\mathbf{x}^{(4)}) \, \mathbf{A}^H}_2^2 %+ \gamma \|\mathbf{x}\|_1
\label{eq:optim}
\end{equation}
% \noindent where the $\ell_2$-norm is 
% minimized to reconstruct $\mathbf{C}$ via Eq. \ref{eq:steer}. %and the $\ell_1$-norm is applied to $\mathbf{x}$ to induce sparsity via the scalar hyper-parameter $\gamma$. 
Note that Eq. \ref{eq:optim} is well-known in compressed sensing \cite{van2013signal, simeoni2019deepwave, chardon2021theoretical}. %By solving this pretext task, a model can be self-supervised to produce a SAM $\mathbf{x}$ by .

% \subsection{The Latent Acoustic Mapping (LAM) model}

% As seen in the previous section, LAM is designed to learn the optimal representation of the sparse vector $\mathbf{x}$ in a self-supervised manner. 

% The network (see Fig. \ref{fig:LAM}) takes as input a CSM $\mathbf{C}$ and
% encodes it %$\mathbf{C}$ 
% via the linear operation %(i.e back-projection) 

% \begin{equation}
% [\bar{\textbf{B}}\circ\textbf{B}]^H \text{vec}(\mathbf{C}) = \text{diag}(\textbf{B}^H\mathbf{C}\textbf{B})
% \label{eq:optim}
% \end{equation}
 
% where $\textbf{B} \in \mathbb{C}^{M\times N}$ is a learnable matrix. This operation, described as a ``back-projection'' in \cite{simeoni2019deepwave, van2013signal}, maps $\mathbf{C}$ to the latent space $\mathbf{x}$. Denoising is performed on $\mathbf{x}$ using four 1D convolutional layers with progressively larger kernels (sizes 3, 5, 7, 9).
% The model's decoder is Eq. \ref{eq:steer}. Using a encoder-decoder architecture, LAM solves Eq. \ref{eq:optim}, finding an optimal SAM $\mathbf{\mathbf{x}}$.

Our training loss function entails: (a) the reconstruction loss, (b) a dispersion term applying a $\ell_1$ regularization to the latent space to make it sparse, %ion strength is 
controlled via $\gamma > 0$, and (c) a total variation regularization \cite{chambolle2010introduction} across neighboring dimensions in the latent space $\mathbf{x}^{(4)}$. The loss function is therefore:

\begin{equation}
L(\mathbf{C}, \mathbf{\hat{C}}) = \text{MSE}(\mathbf{C}, \mathbf{\hat{C}}) + \gamma \big(\norm{\mathbf{x}^{(4)}}_1 + \sum_{i, j\in\mathcal{N}_{i}} |x_i^{(4)} - x_j^{(4)}|\big)
\end{equation}
\noindent It is worth noting that LAM is very lightweight, with the base model having only 16K parameters per frequency band.

\subsection{CSM upsampling for high-resolution SAMs}

\noindent LAM is designed to work with any number of microphone channels $M$. %However, the acoustic map resolution is proportional to the number channels. 
In this study we asses LAM with $M=32$, corresponding to the use of a high-resolution microphone array such as the Eigenmike \cite{acoustics2013em32}. 
However, noting that lower-resolution microphones, such as 4-channel tetrahedral, are more commonly used in SELD datasets \cite{Shimada2023starss23},
we use a Complex-valued Deep Back Projection Network (CDBPN) capable of upsampling $\mathbf{C} \in \mathbb{C}^{4\times 4} \rightarrow \mathbb{C}^{32\times 32}$  \cite{roman2024robust, haris2018deep}. 
This allows LAM to process both 32ch (spherical) and 4ch (tetrahedral) arrays. 

\section{Methodology}
\label{sec:methodology}

\subsection{Datasets}

% Our framework for building LAM utilizes three datasets, with details on their use for cross-validation covered in Sections \ref{sec:self_training} and \ref{sec:supervised_training}.

\underline{Eigenscape unlabeled dataset}: We use the full Eigenscape dataset to carry out the self-supervised training of our model \cite{green2017eigenscape}. 
It encompasses 10 hours of real recordings across 8 acoustic scenes: Beach, Park, Busy Street, Pedestrian Zone, Quiet Street, Shopping Center, Train Station and Woodland.
It was recorded using a 32-channel Eigenmike.

\underline{Simulated datasets}: we use SpatialScaper \cite{roman2024spatial}, which relies on room impulse response (RIR) databases to simulate multichannel sound scapes with spatialized and moving sound events. 
We use 5 RIR databases: METU \cite{olgun2019metu}, ARNI \cite{mckenzie2021dataset}, DAGA \cite{schneiderwind2019data}, RSoANU \cite{chesworth2024room}, and MOTUS \cite{gotz2021dataset}, all of which also feature a 32-channel Eigenmike. 
We generate 2 hours of data per room, hence adding a total of 10 more hours of data.
The simulated sound events were randomly drawn from male and female speech in RAVDESS \cite{livingstone2018ryerson} and FSD50K \cite{fonseca2021fsd50k}, as well as music tracks from FMA \cite{benzi2016fma}. 
This simulated data includes annotations of sound source locations in time and space.%, which we enabling supervised model training. 

\underline{Real labeled datasets}: To evaluate DoAE performance with real recordings, we use the LOCATA and STARSS datasets.
LOCATA was recorded with a 32ch Eigenmike in a room where human actors moved while speaking \cite{lollmann2018locata}. Actors' locations were tracked for ground truth DoA with respect to the microphone. 
STARSS \cite{Shimada2023starss23} features a 4-channel tetrahedral microphone format, recorded and annotated similarly to LOCATA. STARSS is a larger dataset and contains more diversity of sound categories and rooms.

\subsection{Self-supervised training procedure}
\label{sec:self_training}

\noindent We train two variants of LAM, one using 32-channel data that we refer to as LAM, and another one with 4-channel data, which we refer to as UpLAM, as it integrates CBDPN upsampling.
All LAM variants operate on $F=9$ frequency bands linearly spaced from 1.5kHz to 4.5kHz and processing %. LAM's architecture processes 
each band's $\mathbf{C}$ in parallel, ultimately generating $F$ SAMs, one per band.

Both LAM and UpLAM are trained using Eigenscape + simulated data via self-supervised learning (no labels used in this stage).
Tracks are randomly split into training (80\%), and validation (20\%) for optimal model selection via cross-validation. These datasets use the 32 channel Eigenmike, and to obtain the corresponding lower-resolution array (tetrahedral 4ch) we use channels 6, 10, 22, 26, following the approach proposed by \cite{adavanne2019localization}. 
Models are trained with an Adam optimizer \cite{kingma2014adam} with learning rate $lr=1\times 10^{-6}$, and $\gamma=1\times 10^{-4}$. 
LAM is trained with a batch size of 32, while UpLAM is trained with a smaller batch size of 8 to combat overfitting. 

\subsection{LAM for Direction of Arrival Estimation}

\noindent After self-supervised training, we use LAM as a feature extractor for the downstream DoAE task using two approaches: 

\underline{1. K-means}: We stack the different SAMs obtained for each frequency band onto a 3D tensor, resulting in a multi-frequency SAM $I(\textbf{x}) \in \mathbb{R}^{F\times A\times E}$, where A: azimuth \& E: elevation. 
We run weighted K-means clustering with
$K = 3$ on the 18 pixel points with the maximal intensity (all others are clipped to zero).
This approximates three centroids for potential sound locations. 
We then apply an iterative rule for DoAE via K-means: (1) neighboring clusters with centroids within $15^\circ$ of each other are merged together, and (2) centroids separated by more than $15^\circ$ are considered to be independent sound sources.
The K-means parameters ($K$, pixel points, and unification rule) were selected through cross-validation on the validation data. %set used to self-supervise LAM training. %as part of the validation step. 

\underline{2. GRU-MHSA}: After self-supervised learning, we use SAMs as input to a supervised architecture for DoAE.
The goal is to show that SAMs contain optimal information to learn DoAE in a supervised fashion.
We train a model containing two gated recurrent unit (GRU) blocks \cite{chung2014empirical}, followed by two multi-headed self-attention layers (MHSA \cite{sudarsanam2021assessment}). The model is supervised to have the multi-track activity-coupled cartesian DOA (multi-ACCDOA) output representation \cite{shimada2022multi} to track the DoA of multiple and overlapping sound events. 

\subsection{Baseline Models}

Our baseline model for DoAE is SELDnet \cite{adavanne2019localization}. We also evaluarte EINV2 \cite{hu2022track}, which is a higher-performing network. Additionally, we evaluate performance by the MUSIC \cite{schmidt1986multiple} algorithm for high-resolution SAM generation. DeepWave \cite{simeoni2019deepwave} represents a baseline for data-driven high-resolution imaging and also a compressed sensing method considering that its outputs are virtually equivalent to those by top-performing variants of PGD \cite{simeoni2019deepwave}.

\subsection{Supervised DoAE training procedure}
\label{sec:supervised_training}

The 32-channel models for DoAE are trained with an 80:20 random split of data simulated using SpatialScaper \cite{roman2024spatial}. 
The 4ch models are trained using the ``dev-train-tau'' and ``dev-train-sony'' files in the STARSS dataset \cite{Shimada2023starss23} with the companion simulated data for STARSS \cite{politis2020overview}, and validated using ``dev-test-tau''.
For the LOCATA benchmark, models are evaluated on LOCATA tasks 1, 2, 3 \& 4.
For the STARSS benchmark, 4ch models are evaluated using ``dev-test-sony''. 

\subsection{Metrics}

% \noindent \underline{Self-supervision}: During the self-supervised training we monitor the quality of LAM's SAM generation via Peak signal-to-noise ratio (PSNR) on the validation split. We use structural similarity (SSIM) of LAM's SAMs compared to pre-computed DeepWave labels and calculate this metric on the validation split (i.e. not used to update LAM's weights).

\noindent We use Localization Error (LE), which measures the radial difference between predicted and true location per detected sound event, and Localization Recall (LR), which quantifies the number of localized sound events out of the total that are annotated \cite{politis2020overview}. Note that a lower LE indicates higher precision in localization, while a higher LR reflects broader detection coverage.

\section{Results}
\label{sec:results}

Figure \ref{fig:intensity_maps} shows representative SAMs featuring 
two active sources in the LOCATA dataset. The self-supervised LAM and UpLAM are contrasted against the supervised DeepWave. 

Table \ref{tab:locata_results} presents the DoA performance on the LOCATA evaluation, comparing LAM variants against the supervised models of SELDnet \cite{adavanne2019localization, Shimada2023starss23}, EINV2 \cite{hu2022track}, and DeepWave. Our benchmarking shows that compared to other methods LAM can lead to substantial improvements in LE performance across datasets. Notably, LAM consistently reduces LE by more than 40$^o$ relative to MUSIC. When it comes to models using high-resolution 32-channel inputs, LAM$\rightarrow$K-means virtually matches Deepwave in LE and LR, all while being completely self-supervised. The version with the supervised GRU-MHSA head (LAM$\rightarrow$GRU-MHSA) further improves in LE.
These results show that our self-supervised strategy to generate high-resolution SAMs can lead to the best performing model for DoAE. 

\begin{table}
 \begin{center}
 \setlength{\tabcolsep}{3pt}
 \renewcommand{\arraystretch}{1.2} % Adjust row height if needed
 \scriptsize % Reduce font size
 \resizebox{0.75\columnwidth}{!}{
 \begin{tabular}{llll}
   \toprule\toprule
    Input & \textbf{Model} &  $LE$ $\downarrow$ &  $LR$ $\uparrow$ \\ 
   \midrule\midrule
    \multirow{4}{*}{32ch} & MUSIC$\rightarrow$K-means & 56.95$^o$  & 97.2 \\\cline{2-4}& 
    DeepWave$\rightarrow$ K-means\cite{roman2024robust}  & 14.8$^o$  & \bf{99.20}    \\\cline{2-4}    
    & LAM$\rightarrow$K-means & \underline{13.69}$^o$ & 94.0 \\ \cline{2-4}
    & LAM$\rightarrow$GRU-MHSA & \bf{13.41$^o$} &  \underline{94.7} \\ \hline\hline
    \multirow{4}{*}{4ch} & MUSIC$\rightarrow$K-means & 61.83$^o$  & 94.4 \\\cline{2-4}& DeepWave$\rightarrow$K-means\cite{roman2024robust} & 27.10$^o$  & \bf{99.20}    \\ \cline{2-4}
    & UpLAM$\rightarrow$K-means & 23.48$^o$  & \underline{97.8}    \\ \cline{2-4}
    & UpLAM$\rightarrow$GRU-MHSA & \bf{14.44$^o$} &  84.26 \\ 
     \hline\hline
    4ch & SELDnet\cite{adavanne2019localization} & \underline{16.8}$^o$  & 77.96    \\ \hline
    FOA & SELDnet\cite{adavanne2019localization} &  22.43$^o$ & 80.7   \\ \hline
    FOA+4ch & EINV2\cite{hu2022track} & 19.83$^o$  & 80.6    \\ \hline
   \bottomrule
   \end{tabular}}
\end{center}
 \caption{Evaluation of LAM, DeepWave, SELDnet, MUSIC \& EINV2 on LOCATA. Bold numbers indicate the best score among models with 32ch and 4ch inputs. Models whose output is supervised with data are trained three times and the score reflects the average across runs. Boldface and underlined numbers indicate best and second-best scores, respectively.}
 \label{tab:locata_results}
\end{table}

Using lower-resolution 4-channel inputs our model is also competitive. For both the LOCATA (Table \ref{tab:locata_results}) and STARSS (Table \ref{tab:starss_results}) datasets, UpLAM (4ch input) outperforms DeepWave in LE with both K-means post-processing and the supervised GRU-MHSA head, but % for both self-supervised and supervised variants, 
%achieving a $13^\circ$ improvement with the UpLAM$\rightarrow$GRU-MHSA configuration. 
% However, %similar to the 32-channel case, 
its LR deteriorates. % with K-means and with GRU-MHSA. 
Remarkably, UpLAM$\rightarrow$GRU-MHSA surpasses the other supervised approaches of SELDnet (independent of 4ch or FOA input) and EINV2 in both LE and LR on LOCATA, and in LE on STARSS. %, including SELDNet  and  (with FOA+4ch inputs). 
These findings highlight that for low-dimensional microphone arrays, UpLAM can outperforms supervised DoAE methods. % and high-performing deep learning models in terms of LE. The results demonstrate that LAM is a robust and competitive alternative to mainstream audio representations for DoAE.

\begin{figure}
% \centering
% \resizebox{\columnwidth}{!}{
\includegraphics[width=9cm]{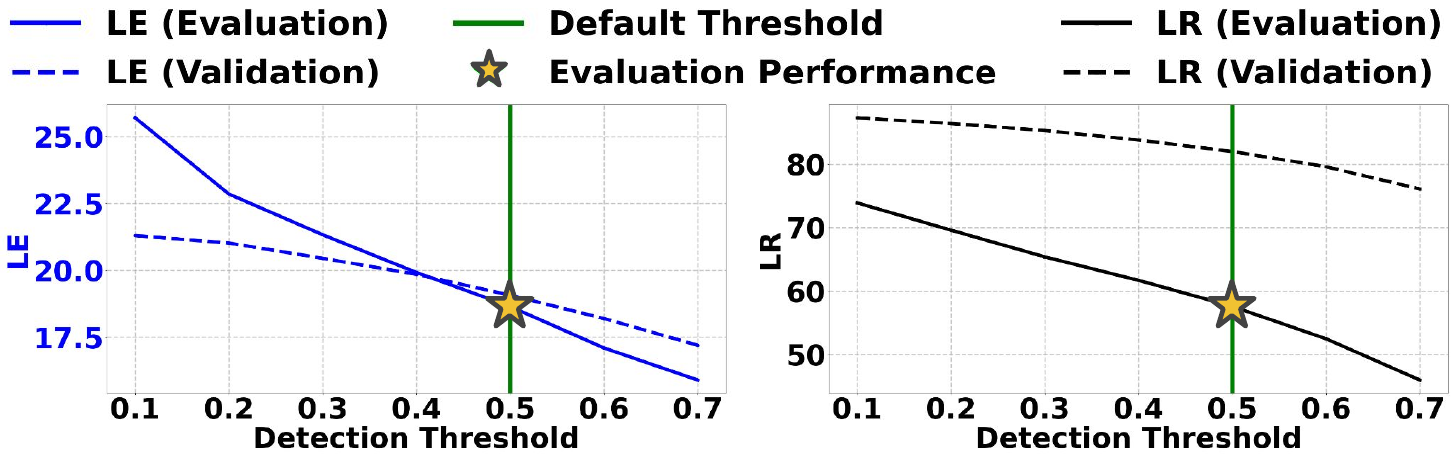}
\caption{UpLAM$\rightarrow$GRU-MHSA LE and LR trade-off as a function of the Multi-ACCDOA detection threshold used at inference. The green vertical line shows model performance at the default threshold of 0.5.}
\label{fig:detection_thresh}
\end{figure}

Why is the LR metric lower when when combining UpLAM with the GRU-MHSA classifier? This is particularly notable in the STARSS dataset. Figure \ref{fig:detection_thresh} provides further insight on how LE and LR are modulated by the detection threshold in the Multi-ACCDOA output, which is use to binarize the activity of sound events \cite{shimada2022multi}. The third row in Table \ref{tab:starss_results} corresponds to test set performance using the default threshold of 0.5 (green vertical line). Reducing the threshold makes the model more sensitive to detect sound events (higher LR) but at the expense of worse LE. Interestingly, the LE deteriorates faster on the evaluation set (blue solid line) as a function of the reduced threshold, while the LR for both validation (black dotted line) and evaluation (black solid line) sets improve at a similar rate. 
Note that around 0.5, the validation and evaluation LE are similar, but the LR is much lower in the evaluation.
%In addition, for thresholds below 0.5, 
We hypothesize that this result highlights a problem not necessarily with UpLAM's SAM but with the GRU-MHSA classifier being susceptible to overfitting due to the small size of STARSS. In other words, the specific sounds and acoustic from the ``dev-test-sony'' evaluation set represent a domain-shift for the model.

\begin{table}
 \begin{center}
 \setlength{\tabcolsep}{3pt}
 \renewcommand{\arraystretch}{1.1} % Adjust row height if needed
 \scriptsize % Reduce font size
 \resizebox{0.75\columnwidth}{!}{
 \begin{tabular}{llcc}
   \toprule\toprule
    \textbf{Model} & Input &  $LE$ $\downarrow$ &  $LR$ $\uparrow$ \\ 
   \midrule\midrule    
    MUSIC$\rightarrow$GRU-MHSA & 4ch & 39.9$^o$  & 73.5  \\ \hline
    DeepWave$\rightarrow$GRU \cite{roman2024robust} & 4ch & 20.5$^o$ &  70.1 \\ \hline    
    UpLAM$\rightarrow$GRU-MHSA & 4ch & \underline{18.65$^o$} &  57.6 \\ 
    \hline\hline
    SELDnet \cite{adavanne2019localization} & 4ch & 23.3$^o$  & 82.3    \\ \hline
    SELDnet \cite{adavanne2019localization} & FOA & 21.9$^o$  & \underline{83.0}   \\ \hline
    EINV2 \cite{hu2022track} & FOA+4ch & 24.0$^o$  & \bf{84.2}   \\ 
    \hline\hline
    UpLAM$\rightarrow$GRU-MHSA $\dagger$ & 4ch & \bf{18.31$^o$} &  64.6 \\ \hline
    SELDnet \cite{adavanne2019localization} $\dagger$ & 4ch & 20.8$^o$  & 66.4    \\ \hline
    \bottomrule
   \end{tabular}}
\end{center}
 \caption{Performance of LAM, DeepWave, SELDnet, MUSIC \& EINV2 on the STARSS ``dev-test-sony'' split (held out for evaluation). Boldface and underlined numbers indicate best and second-best scores, respectively. $\dagger$ denotes experiments where LOCATA \& RSoANU were included in the validation set.}
 \label{tab:starss_results}
\end{table}

To assess this, we enriched the validation set to also include all of LOCATA (real-world recordings) and RSOANU (simulated data). We also ran this experiment with SELDnet as an experimental control. The two bottom rows of Table \ref{tab:starss_results} show these results. Both models saw slight improvement in LE, but UpLAM$\rightarrow$GRU-MHSA saw a significant improvement in LR performance. This indicates that a domain-shift effect is indeed at play, since enrichment of the validation set combated overfitting for the LR metric. It also suggests that UpLAM does indeed contain the information to do the DoAE task, since in these experiments UpLAM was frozen and only the GRU-MHSA was trained. In other words, the supervised head on top of UpLAM is the one lacking generalization, and not UpLAM itself.

\section{Conclusion}
\label{sec:conclusion}

We presented LAM, a self-supervised model for high-resolution acoustic imaging. LAM demonstrates robustness from learning spatial audio representations of large data and enables adaptability to process both low- and high-resolution microphone arrays. Our evaluation shows that it is possible to attain competitive localization performance on the LOCATA dataset using an the fully self-supervised LAM with K-means post processing approach, surpassing MUSIC and DeepWave. %, SELDnet and EINV2. 
Additionally, when LAM's SAMs are used as features for supervised models, localization performance on STARSS and LOCATA datasets exceeds that of all other methods, highlighting robustness and versatility across real-world scenarios. %However, these gains come with LR performance generalization trade-offs, which can be addressed by simply using a validation set that contains rich acoustic diversity.

LAM reduces reliance on annotations for deep acoustic imaging, demonstrating versatility in supervised and unsupervised DoAE tasks. Future work aims to extend its use to sound event classification and other signal processing domains like antenna arrays. With only 16K parameters (significantly fewer than SELD deep neural networks with over a million), its compact architecture makes it well-suited for real-time processing and deployment on edge devices.

\clearpage

\section{Acknowledgments}
\noindent This work was partially supported by the National Science Foundation grant no. IIS-1955357. The authors thank the funding source
and their grant collaborators.

% -------------------------------------------------------------------------
% Either list references using the bibliography style file IEEEtran.bst

 % . \\
% The \IEEEtriggeratref{XX} command can be used to move to the next column before the XX-th reference
% to balance the two columns of the reference section
% \IEEEtriggeratref{XX}
\bibliographystyle{IEEEtran}
\bibliography{refs25}
% or list them by yourself:
% \begin{thebibliography}{1}

% \bibitem{waspaaweb}
% {WASPAA Website}, \url{http://www.waspaa.com}.

% \bibitem{IEEEXploreReqs}
% {IEEE {X}plore {R}equirements}, \url{https://conferences.ieeeauthorcenter.ieee.org/write-your-paper/meet-ieee-xplore-requirements/}.

% \bibitem{eWilliams1999}
% E.~Williams, \emph{Fourier Acoustics: Sound Radiation and Nearfield Acoustic Holography}.\hskip 1em plus 0.5em minus 0.4em\relax London, UK: Academic Press, 1999.

% \bibitem{cJones2003}
% C.~Jones, A.~Smith, and E.~Roberts, ``A sample paper in conference proceedings,'' in \emph{Proc. ICASSP}, vol.~II, Apr. 2003, pp. 803--806.

% \bibitem{aSmith2000}
% A.~Smith, C.~Jones, and E.~Roberts, ``A sample paper in journals,'' \emph{IEEE Trans. Signal Process.}, vol.~62, pp. 291--294, Jan. 2000.

% \end{thebibliography}

\end{document}